\documentclass[twocolumn,showpacs,showpacs,superscriptaddress]{revtex4-1}
\usepackage{graphicx,amsmath,amssymb,appendix}
\usepackage[colorlinks,linkcolor=blue,citecolor=blue,urlcolor=blue]{hyperref}
\usepackage{color}

\begin{document}
\title{Boosting cooperation by involving extortion in spatial Prisoner's dilemma}
\author{Zhi-Xi Wu}\email{eric0724@gmail.com}
\affiliation{Institute of Computational Physics and Complex Systems, Lanzhou University, Lanzhou Gansu 730000, China}
\author{Zhihai Rong}\email{zhihai.rong@gmail.com}
\affiliation{Web Sciences Center, University of Electronic Science and Technology of China, Chengdu Sichuan 611731, China}
\affiliation{Department of Electronic and Information Engineering, The Hong Kong Polytechnic University, Kowloon, Hong Kong}

\date{Received: date / Revised version: date}
\begin{abstract}
We study the evolution of cooperation in spatial Prisoner's dilemma games with and without extortion by adopting aspiration-driven strategy updating rule. We focus explicitly on how the strategy updating manner (whether synchronous or asynchronous) and also the introduction of extortion strategy affect the collective outcome of the games. By means of Monte Carlo (MC) simulations as well as dynamical cluster techniques, we find that the involvement of extortioners facilitates the boom of cooperators in the population (and whom can always dominate the population if the temptation to defect is not too large) for both synchronous and asynchronous strategy updating, in stark contrast to the otherwise case, where cooperation is promoted for intermediate aspiration level with synchronous strategy updating, but is remarkably inhibited if the strategy updating is implemented asynchronously. We explain the results by configurational analysis and find that the presence of extortion leads to the checkerboard-like ordering of cooperators and extortioners, which enable cooperators to prevail in the population with both strategy updating manners. Moreover, extortion itself is evolutionary stable, and therefore acts as the incubator for the evolution of cooperation.

\end{abstract}
\pacs{89.75.Fb, 87.23.Kg, 02.50.Le, 89.65.-s}
\maketitle

\section{Introduction}
Cooperation means doing good for others incurring a cost to oneself, which is indispensable for the organization and functioning of societies from insect to human~\cite{Nowak2006book}. Yet, the emergence and persistence of cooperation is an evolutionary riddle as it defies the basic principles of Darwinian evolution theory~\cite{Sigmund2010book}: If cooperation is costly to the individual and benefits only the interaction partners, then Darwinian selection should favor non-cooperating defectors and eliminate cooperation. Over the past decades, the evolution of cooperation has attracted considerable attention in scientific communities, ranging from sociologists, biologists, mathematicians, to physicists~\cite{Axelrod1981science,Axelrod2006book,
Smith1982book,Hofbauer1998book,Doebeli2005el,Nowak2006science,
Rand2013tcs,Szabo2007pr,Roca2009review,Perc2010bio, Ohtsuki2006nature,Gomez2007prl}.

Perhaps, the most famous metaphor for cooperation is the prisoner's dilemma (PD) game, where two confronting players can choose either to cooperate or to defect. Mutual cooperation yields the reward $R$, mutual defection leads to punishment $P$, and the conflicting choice gives the cooperator the sucker's payoff $S$ and the defector the temptation $T$. The parameters satisfy $T>R>P>S$ so that without extra incentives to support cooperation, the defector will always beat the opponent~\cite{Axelrod2006book}. Besides the well known five rules (kin selection, direct and indirect reciprocity, spatial reciprocity, and group selection) reviewed in~\cite{Nowak2006science} favoring the evolution of cooperation, recently discovered prominent mechanisms fostering cooperation are heterogeneous interaction structure~\cite{Santos2005prl}, dynamical linking~\cite{Pacheco2006prl}, distinct interaction and learning graphs~\cite{Ohtsuki2007prl,Wu2007pre}, partner switching~\cite{Fu2009pre}, diversity of impact weight and time scales~\cite{Yang2009pre,Wu2009pre,Rong2010pre,Rong2013epl}, dynamic payoff matrix~\cite{Lee2011prl}, migration~\cite{Roca2011pnas,Yang2010pre,Chen2012pre}, interdependent interaction network~\cite{Wang2013sr}, and payoff sharing~\cite{Wu2014pre}, to name just a few. We refer to Refs.~\cite{Nowak2006science,Rand2013tcs,Szabo2007pr,Roca2009review,
Perc2010bio} for brief overview on the state of the art.

Recently, Press and Dyson showed a novel class of strategies via two-person repeated PD game, so-called zero-determinant (ZD) strategies, which allow players to enforce a linear relation unilaterally between one player's own payoff and the coplayer's payoff~\cite{Press2012pnas}. Especially, a subset of ZD strategies, extortion strategies, have attracted considerable attention~\cite{Press2012pnas,Stewart2012pnas,Stewart2013pnas, Adami2013nc,Hilbe2013pnas,Szolnoki2014pre}. Since extortioners can guarantee that one player's own surplus exceeds the coplayer's surplus by a fixed percentage, extortion is therefore able to dominate any evolutionary opponent~\cite{Press2012pnas}. In the realm of evolutionary games, however, the success of strategies is determined not only by their performance when confronting against other opponents, but also by the performance when confronting against themselves. Actually, in a homogeneous population of extortioners, it is better to deviate by cooperating, since any two extortioners hold each other down to surplus zero, and extortion is therefore evolutionarily unstable in well-mixed population~\cite{Adami2013nc,Hilbe2013pnas}.

The evolutionary viability of extortion in structured populations has been discussed recently by Szolnoki and Perc~\cite{Szolnoki2014pre, Szolnoki2014sr}. It was found that extortion is evolutionary stable if the strategy updating is governed by a myopic best response rule. The spirit of the myopic best response rule is that if we are not happy or satisfied with the benefit reaped by adopting the current strategy, we may explore something else. In fact, myopic best response rule is one of the heuristic strategies based on stochastic learning theory~\cite{Macy2002pnas}. Other classical examples include tit-for-tat, win-stay-lose-shift~\cite{Nowak2006book}, reinforcement learning~\cite{Szabo2007pr}, etc. Indeed, learning theory has provided a powerful and reasonable framework to study the evolution of cooperation, since players can be viewed as adaptive agents, who are able to responding to and learning from the ever-changing environment in perpetuity.

In this work, we continue the research line of~\cite{Szolnoki2014pre} to explore how the incorporation of extortion in the PD affects the evolution of cooperation and the evolutionary viability of extortion as well in structured populations, but instead of the myopic best response rule, we focus on games with aspiration-driven strategy updating rule~\cite{Chen2009pre}. In Ref.~\cite{Chen2009pre}, Chen and Wang studied the evolution of cooperation in the context of weak PD (where $R=1$, $T=b$, and $S=P=0$) with a stochastic learning rule~\cite{Macy2002pnas}, where players update their strategies depending on the difference between the actual and aspiration payoffs, and found that some certain moderate aspiration level can lead to the optimal cooperation level.

We note that the strategy updating of the players is executed synchronously in their model. It has been pointed out that whether the strategy updating is implemented synchronously or asynchronously usually leads to large disparities in the final evolutionary outcome~\cite{Hubermann1993pnas}. We will show below that the qualitative properties of the results reported in~\cite{Chen2009pre} would change dramatically given that an asynchronous strategy updating scheme is instead employed. In particular, intermediate aspirations will give rise to lowest cooperation level. Unexpectedly, the qualitative disparities induced by different strategy updating manners will go to disappear if extortion strategy is introduced into the games, which highlights the positive role of extortion in boosting and sustaining cooperation.

\section{Model}
Most actual populations are spatially structured in the sense
that individuals usually interact with those who share close
geographic proximity. To account for this feature, we
consider evolutionary games in spatially structured
population, where the players are located on the sites of a
square lattice with periodic boundary conditions. Each player can take one of the three competing strategies, namely extortion $E_{\chi}$, cooperation $C$, and defection $D$. Following closely with the previous work~\cite{Hilbe2013pnas,Szolnoki2014pre,Szolnoki2014sr}, we will account for the popular form of the true PD, i.e., the so called donation game, in which cooperators provide a benefit $b$ to their partners at a cost $c$ to themselves ($b>c>0$), while defectors provide neither benefits nor incur costs. With extortioners participating in the game, the payoff matrix now reads as
\begin{equation}\label{payoff}
\begin{tabular}{r|c c c}
 & $E_\chi$ & $C$ & $D$ \\
\hline
$E_\chi$ & 0 & $\frac{(b^2-c^2)\chi}{b\chi + c}$ & 0 \\
$C$ & $\frac{b^2-c^2}{b\chi+c}$ & $b-c$ & $-c$ \\
$D$ & 0 & $b$ & 0 \\
\end{tabular}
\end{equation}
where $\chi$ determines the surplus of the extortioner in relation to the surplus of the other player~\cite{Press2012pnas,Stewart2012pnas,Stewart2013pnas, Adami2013nc,Hilbe2013pnas}. Note that with the absence of the extortion strategy, the donation game is no other than a true PD game. The parameter $b$ determines the strength of the social dilemma, while the parameter $\chi$ determines just how strongly the strategy $E_\chi$ exploits cooperators. For simplicity, just as in~\cite{Szolnoki2014pre}, we set $\chi=1.5$ and $b-c=1$ so that there is only one free parameter in the payoff matrix, i.e., the cost to cooperate $c$. We have validated that our main conclusions presented below are robust to other choice of large $\chi$ (e.g., $\chi=5$).

Initially, each player holds one of the three (or two, if extortion is not incorporated into the game dynamics) strategies with equal probability. The players collect gains according to the payoff matrix~(\ref{payoff}) through playing with their immediate neighbors. We consider an aspiration-driven strategy updating to study the evolution of strategies. During the evolutionary processes, the players always try to explore other possible strategies dependent on the difference between their current payoff and the aspiration level, which can be understood as the extent of satisfaction of the players with their environments (for instance, the aspiration level could be regarded as the minimum living standard). Following Chen and Wang~\cite{Chen2009pre}, the aspiration level $P_{xa}$ for one certain player $x$ is defined as $P_{xa}=k_xA$, where $k_x$ is the number of neighbors of $x$ and $A$ the control parameter ($k_x=4$ for the square lattice, and $A$ is the same for all the players). The value of $A$ is typically constrained to the interval [0, $b-c$]. Whenever updating his strategy, a player $x$ will move to another randomly selected possible strategy with a probability
\begin{equation}\label{rule}
q=\frac{1}{1+\exp[(P_x-P_{xa})/\kappa]},
\end{equation}
where $P_x$ is the actual payoff yielded by $x$, and $\kappa$ the noise factor, which characterizes the irrationality or uncertainty of the players in strategy transformation~\cite{Szabo1998pre,Blume1993geb,Traulsen2007jtb}. Herein, $\kappa$ is simply set to 0.1 to practically prohibit a strategy change if the current strategy yields a higher payoff than the aspiration level (we note that the qualitative features remain unchanged for different $\kappa$).

We account for two classical strategy updating schemes for our model: one is the synchronous updating, and the other is the asynchronous (or random sequential) one. In the former case, in each MC time step all the players get payoffs by playing with their neighbors and update their strategies simultaneously according to Eq.~(\ref{rule}), while in the latter case the players are selected by chance to change their strategies, and one MC time step consists of all the players having updated their strategies once on average. Regardless of the applied strategy updating manner, we iterate the above elementary processes for enough long time so that the average frequency of strategies becomes time independent. In all simulations, the stationary frequency of strategies are obtained by averaging over the last $3000$ generations of the entire $10^4$ ones, and the data presented below are obtained by averaging over $20$ independent realizations. All our MC simulations are carried out on a square lattice with size $N=201\times201$.

\section{Results and Analysis}
Let us firstly present the simulation results of our model entailing only strategies $C$ and $D$. Regardless whether the strategy updating is performed synchronously or asynchronously, for too small as well as too large $A$, $f_C$ in the steady state will be around 0.5, which are the expected trivial cases. For sufficiently small $A$, all the players are satisfied with what they yield from their neighbors through game interactions, and the system is actually ``frozen" in its initial state, whereas for high enough $A$, all the players will be always unhappy and just switch their strategies blindly for ever (such ``boiled" system would naturally result in $f_C=0.5$). Thus, what interests us most is the region where $A$ is not too small, and not too large as well. Hereafter, we will restrict our main attention to the region $0<A<b-c=1.0$.

\begin{figure}
\includegraphics[width=\linewidth]{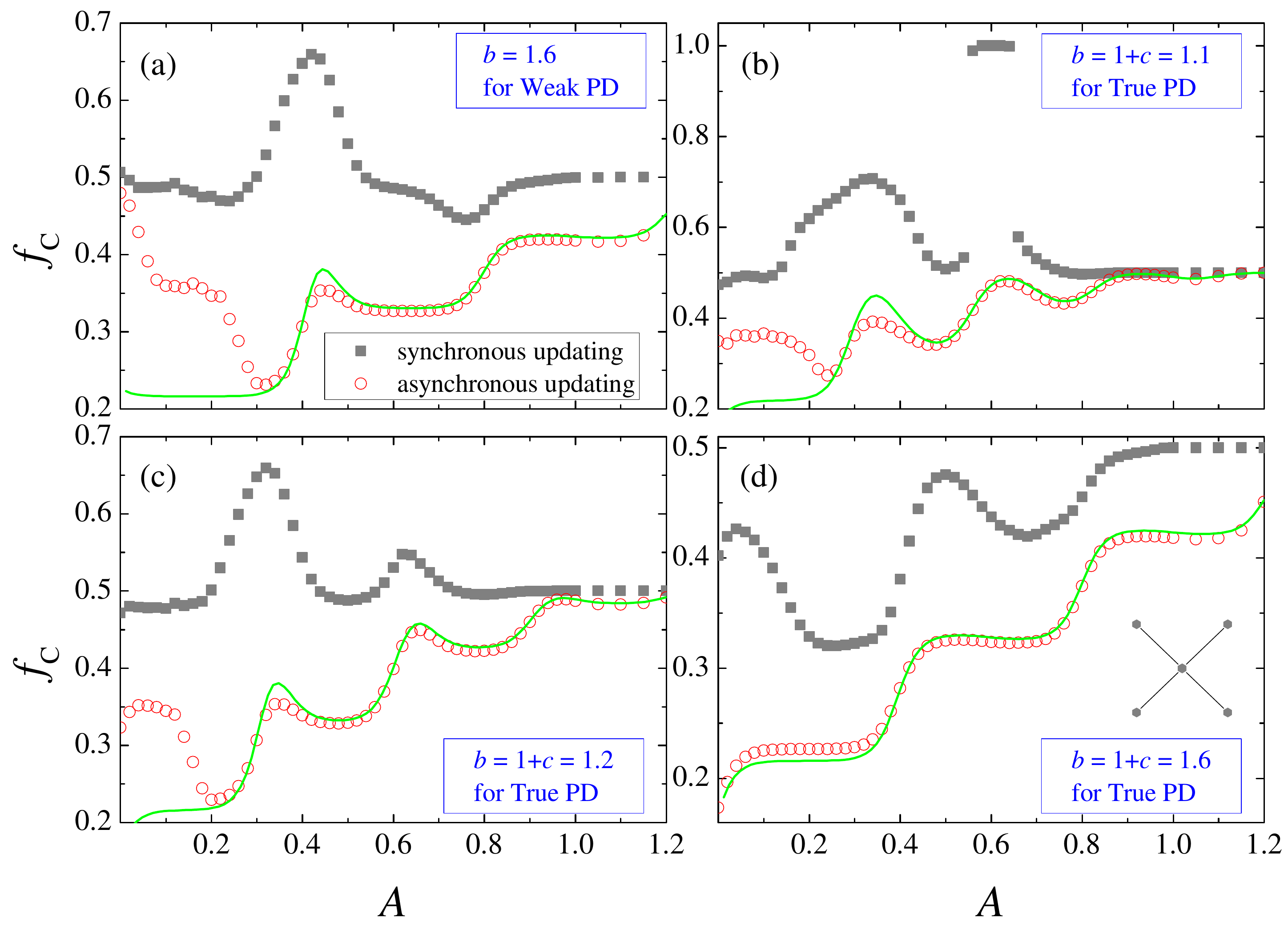}\\
\caption{(Color online) Stationary frequency of cooperation $f_C$ as a function of the average aspiration level $A$ for different values of the temptation to defect $b$. (a) is for the weak PD case with $R=1$, $T=b$, and $S=P=0$; (b)-(d) are for the true PD case with $R=b-c=1$, $T=b$, $S=-c$, and $P=0$. The solid squares and open circles correspond, respectively, to the results obtained by carrying out synchronous and asynchronous strategy updating, and the continuous lines represent the predictions of the five-site [as shown in panel (d)] dynamical cluster approximation~\cite{Vukov2006pre,Szabo2005pre,Szabo1998pre}.}\label{cd}
\end{figure}

The results summarized in Fig.~\ref{cd}(a)-(d) show clearly that intermediate aspiration level can give rise to the enhancement of cooperation for both donation game and weak PD game, provided that the strategy updating is implemented in a synchronous way. Specifically, for not too large $b$ [e.g., $b=1.1$ in Fig.~\ref{cd}(b)], the maximum value of $f_C$ can even go up to 1, that is, cooperators dominate totally the population. Nevertheless, we will get the complete opposite scenario if the strategy updating of the players is performed asynchronously. In particular, we observe that some moderate aspirations result in the lowest cooperation level, and the value of $f_C$ of is always smaller than $0.5$ in the whole aspiration region. Rather than as a promoter of cooperation in the case of synchronous strategy updating, medium aspirations now behave as an inhibitor for the formation and maintenance of cooperation. The generalized mean filed approximation method based on five-site cluster~\cite{Vukov2006pre,Szabo2005pre,Szabo1998pre} predicts correctly the evolutionary trend of cooperation for the asynchronous strategy updating case (Figure~\ref{cd} shows explicitly that the estimations obtained via five-site cluster dynamical mean-field approximations match more and more accurate with the simulation results as $A$ increases, especially for large $b$). It is worthy pointing out that the dynamical cluster approximation is based on the assumption of continuous time, and hence on the asynchronous (random sequential) strategy updating~\cite{Szabo2007pr}.

The discrepancy induced by the different strategy updating manners can be understood intuitively as follows. We know that in the context of donation game, mutual cooperation distributes benefits to each others, while mutual defection does not. Thus, the clusters composed of cooperators will be much more stable than those of defectors as long as $A<b-c=1.0$. In the case of synchronous strategy updating, those clustered, unsatisfied defectors may change to cooperators simultaneously. After that, the clustered cooperators become happy and stable since they benefit each other. As such, the synchronous strategy updating manner could facilitate the formation of large clusters of cooperators for moderate aspiration levels, hence promoting cooperation. In contrast, if the players update their strategies asynchronously, the clusters of defectors (if formed) are stable and there is little incentives for inner defectors to change to cooperation, and occasional strategy flippings will be restored rapidly, since any tentative transformation to cooperators just means that they would be exploited more severely by the defective neighbors in the future. As a result, medium aspirations just deteriorate the circumstance to incubating cooperation.

\begin{figure}
\includegraphics[width=\linewidth]{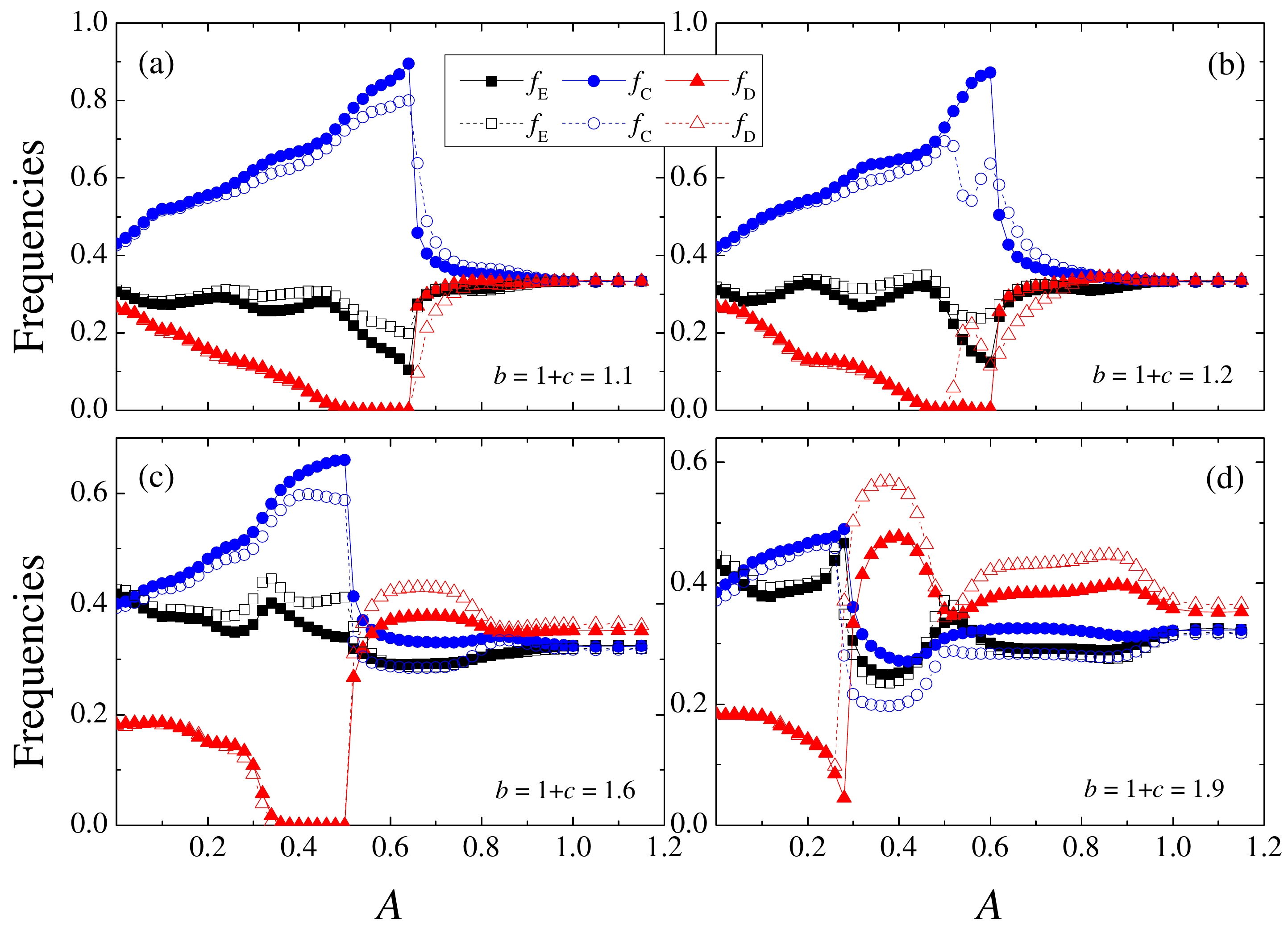}\\
\caption{(Color online) Stationary frequencies of extortion $f_E$ (squares), cooperation $f_C$ (circles), and defection $f_D$ (triangles) as a function of the average aspiration level $A$ for different values of the temptation to defect $b$. Here we have selected $\chi=1.5$ to define the extortion strategy $E_{\chi}$. Other parameters are $R=b-c=1$, $T=b$, $S=-c$, and $P=0$. Solid and hollow symbols correspond, respectively, to the results obtained by carrying out synchronous and asynchronous strategy updating. The lines are guides to the eyes.}\label{ecd}
\end{figure}

The above disagreement between the results yielded by synchronous and asynchronous strategy updating simulates us to investigate whether the involvement of extortioners in the donation game would affect the final evolutionary outcome. In what follows, we report our results for the three-strategy competing game model. In Fig.~\ref{ecd}, the stationary frequencies of the three strategies, $f_E$, $f_C$, and $f_D$, are plotted as a function of $A$ for four special values of $b$. Remarkably, we see that the introduction of extortion can help cooperators to establish and even dominate in the population for appropriate intermediate aspiration level $A$. Contrary to the case of two-strategy donation game, now both the synchronous and asynchronous strategy updating result in the qualitatively same results: cooperation is boosted in a resonance-like behavior for intermediate aspiration level $A$, despite of the fact that the introduction of extortioners increases the chance of cooperators being exploited. Moreover, extortion itself becomes evolutionary stable in the whole range $0<A<b-c=1$ for all the considered values of $b$. Particularly, in the region where cooperation is promoted (that is, $f_C$ in the stationary state is greater its initial value 1/3), extortion always outperforms defection.

Note that for not large values of $b$ [$b=1.1$ and 1.2 in Figs.~\ref{ecd}(a) and (b)], the average fraction of cooperators in the stationary state is always greater than those of extortioners and defectors. Even for sufficiently large $b=1.6$ and $1.9$, cooperators are the majority in the population as long as $A\leq0.5$ and $A\leq0.3$ [Figs.~\ref{ecd}(c) and (d)]. Peculiarly, for moderate aspiration levels, the presence of extortion can even drive the defectors to extinction [see Fig.~\ref{ecd}(a)-(c)]. Hence, unlike the work by Szolnoki and Perc~\cite{Szolnoki2014pre}, where the same three-strategy game model is studied by adopting myopic best response rule and the existence of extortioners is proven to just provide an evolutionary escape hatch for cooperator to survive, in our current model extortion behaves as the incubator and promotor for the emergence and persistence of cooperation.

\begin{figure}
  \includegraphics[width=\linewidth]{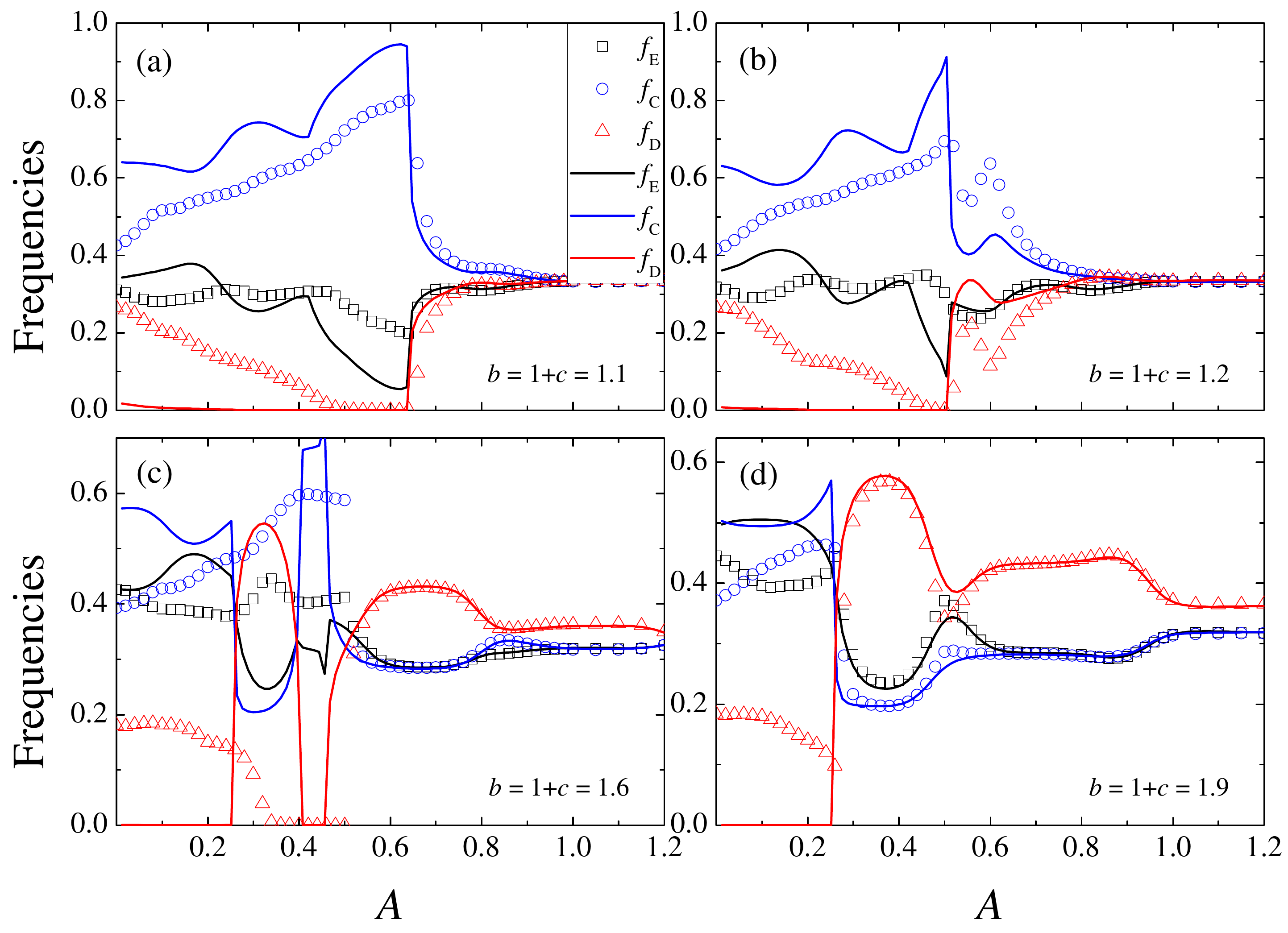}\\
  \caption{Stationary frequencies of extortion $f_E$, cooperation $f_C$, and defection $f_D$ as a function of the average aspiration level $A$ for different values of the temptation to defect $b$. The symbols are exactly the same as shown in Fig.~\ref{ecd} for the case of asynchronous strategy updating, and the solid lines illustrate the results predicted by the five-site dynamical cluster approximation~\cite{Vukov2006pre,Szabo2005pre,Szabo1998pre}.} \label{dynclu}
\end{figure}

For the asynchronous strategy updating case, we are also able to estimate the average frequencies of the three strategies by using of the generalized mean-field approximation method~\cite{Vukov2006pre,Szabo2005pre,Szabo1998pre}. Once again, the predictions of the five-site cluster approximation shown in Fig.~\ref{dynclu} are in good consistent with the MC simulation results, especially for large values of $A$.

\begin{figure}
\includegraphics[width=0.48\linewidth]{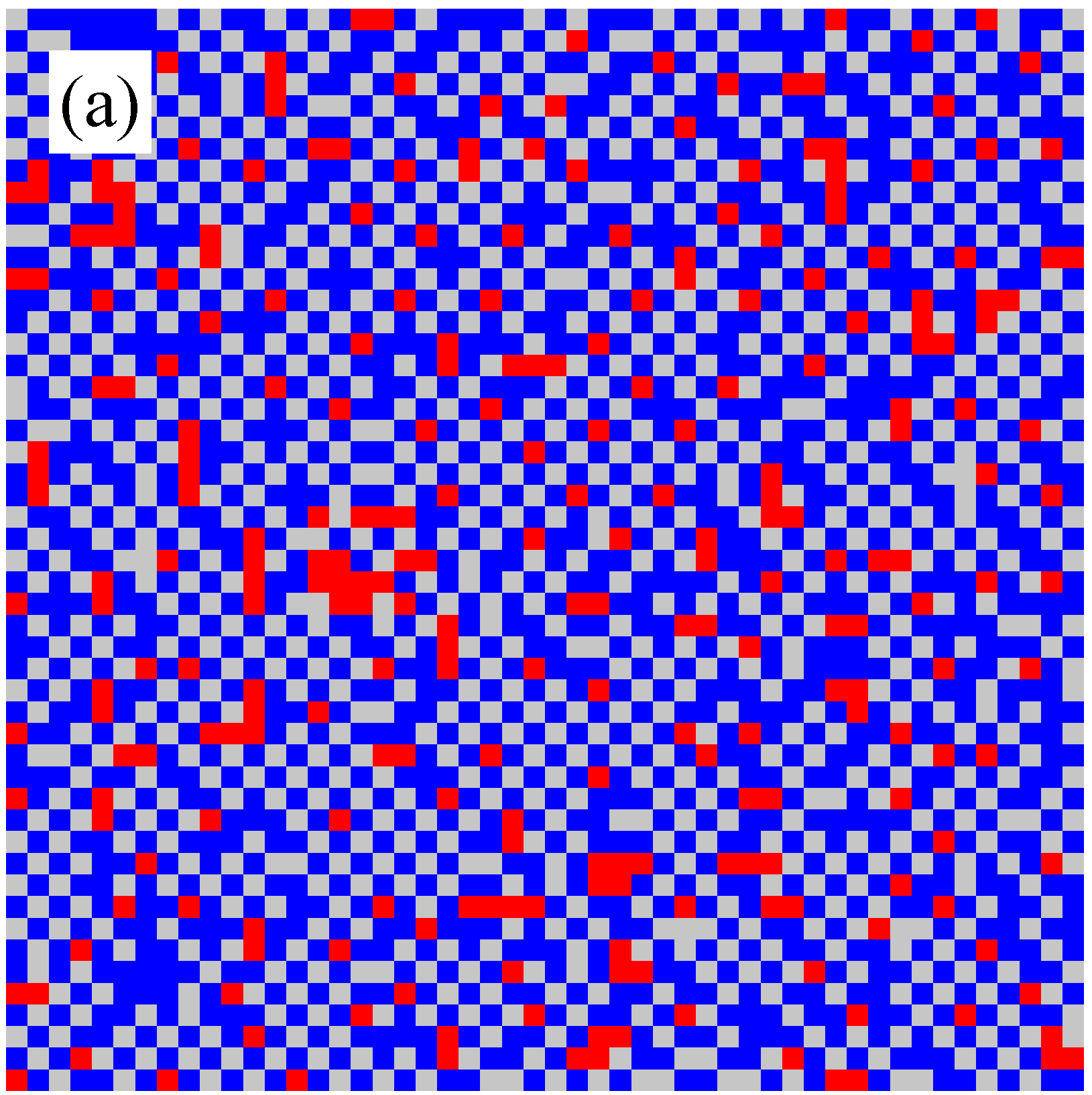}
\includegraphics[width=0.48\linewidth]{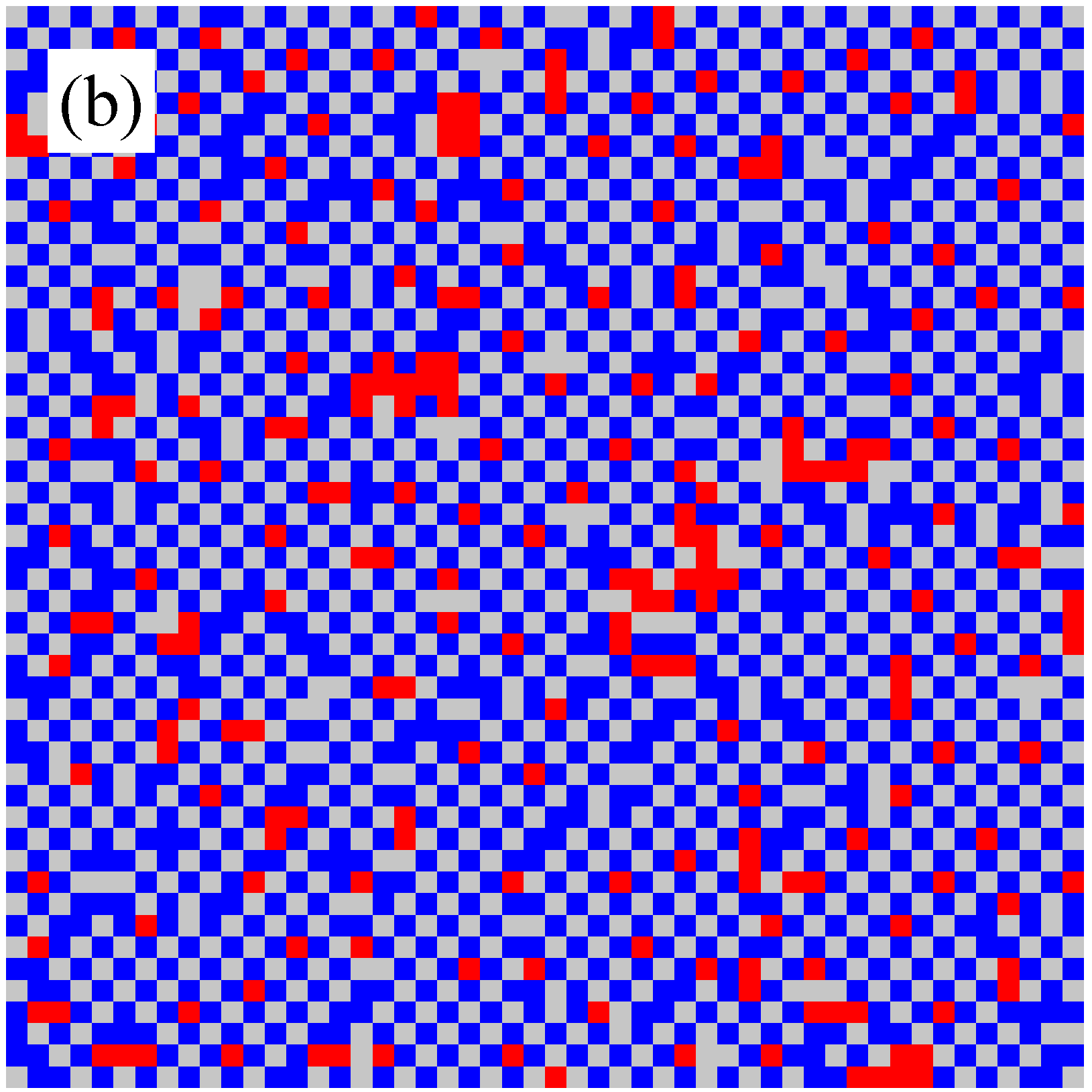}\\
\caption{(Color online) Typical spatial distributions of extortioners (light gray), cooperators [blue (dark gray)], and defectors [red (medium gray)], as obtained for the three-strategy competing model with $A=0.3$ and $c=0.6$ on the square lattice by means of aspiration-driven strategy updating. (a) is for synchronous strategy updating, and (b) for asynchronous strategy updating rule. Other parameters are the same as in Fig.~\ref{ecd}.}\label{config}
\end{figure}

To explain these unexpectedly results, we monitor the typical stationary distribution of strategies for the parameter combination $A$=0.3 and $b$=1+$c$=1.6, under which condition the average frequencies of extortion, cooperation, and defection satisfy the relationship $f_C>f_E>f_D$. The pictures displayed in Figs.~\ref{config}(a) and (b) correspond, respectively, to the cases where the strategy updating is executed synchronously and asynchronously. At a first glance, we observe that cooperators and extortioners self-organize into checkerboard-like configuration, just like what has been found in~\cite{Szolnoki2014pre} where the game dynamics is driven by the myopic best response rule. This is somewhat expected since $E_{\chi}$ and $C$ form the snowdrift-like relation~\cite{Szolnoki2014pre,Szolnoki2014sr,Hilbe2013pnas}.

In order to get a deep insight on how the system evolves to checkerboard-like patterns, let us consider the case $b$=1+$c$=1.2 [Fig.~\ref{dynclu}(b)]. When $A\approx0$, the $C$ players with many $D$ neighbors are unhappy with what they get and would transform to either $D$ or $E_{\chi}$, which will increase the abundance of $D$-$E_{\chi}$ and $E_{\chi}$-$E_{\chi}$ pairs in the system. After that, the $E_{\chi}$ ($D$) players with neighbors comprising only $E$ and/or $D$ begin to get unsatisfied since they get nothing from each others. Driven by the rule of~(\ref{rule}), in the sea of defectors, $D$ will drift to $E_{\chi}$, and once the sea of extortioners is formed by chance, $C$ gets the great chance to strike back and flourish (since a cooperator confronting extortioners will get more than extortioners confronting themselves). Because $C$ and $E_{\chi}$ are able to support each other due to their snowdrift-like relation, the $C$-$E_{\chi}$ pair is fairly stable than the $D$-$E_{\chi}$ and $E_{\chi}$-$E_{\chi}$ pairs. Consequently, the equilibrium $f_C$ will be greater than $f_E$ and $f_D$. With increasing $A$, those $D$ players with just one $C$ neighbor also begin to becoming unsatisfied. Moreover, the $C$-$C$ pairs are more likely to evolve to $C$-$E_{\chi}$ pairs. Taken together, as long as $A$ does not go to too large (where no players are happy with what they get and all the three strategies switch randomly), the abundance of $C$-$E_{\chi}$ pair in the system will grow steadily, while $D$-$D$, $D$-$E_{\chi}$, and $E_{\chi}$-$E_{\chi}$ decrease stably. The running off of $E_{\chi}$ induced by the pairs $E_{\chi}$-$E_{\chi}$ changing to $C$-$E_{\chi}$ could be compensated by the pairs $D$-$E_{\chi}$ changing to $E_{\chi}$-$E_{\chi}$, such that the $f_E$ is nearly unchanged. The ultimate result is that $f_C$ ($f_D$) will grow (decline) with increasing $A$, and cooperators and extortioners self-organize into checkerboard-like pattern, similar to those shown in Fig.~\ref{config}. As $A$ goes to sufficiently large, say $A>b-c=1$, the average frequencies of the three strategies converge definitely to the same value $1/3$.

We notice that for sufficiently high $b$, say $b=1.6$ and $1.9$, defectors will become the majority in the population and the equilibrium $f_D$ will be greater than $f_E$ and $f_C$, if the aspiration level goes beyond a certain \emph{critical} value [see Figs.~\ref{ecd} and~\ref{dynclu}]. The main reason may come from the fact that for large enough $A$, the $C$ players surrounded by two (or even one) $D$ neighbors are becoming unsatisfied (remind that large $b$ also means large cost $c$, since we have fixed $b-c=1$), and they are more likely to transfer to defection, whereas the $D$ players would be happy if they had only one neighbor of $C$. The heavy exploitation on cooperators by $D$ restrains the further growth of $C$.

So far we have witnessed that introducing extortioners into the original donation game will lead to the most possible strategy configuration in the stationary state that all the three strategies are scattered among each others~(Fig.~\ref{config}). Actually, the essential reason for this emergent phenomenon is due to the fact that the strategies $E_{\chi}$ and $D$ are neutral and the strategies $E_{\chi}$ and $C$ are of snowdrift-like relation in terms of~(\ref{payoff}), both of which support the scattering of strategies and prevent the clustering of them. The emergent scattering pattern of strategies can efficiently avoid the players to flip collectively their strategies when the strategy updating is implemented synchronously, in contrast to the case of two-strategy donation game (Fig.~\ref{cd}). Accordingly, both synchronous and asynchronous strategy updating schemes give rise to the qualitatively similar results for the studied three-strategy competing game model, despite of some slight quantitative differences.

Finally, it is worthy pointing out that the checkerboard-like configuration composed of $E_{\chi}$ and $C$, once formed, can efficiently restrain the growth of $D$ for small and intermediate $A$ in the framework of aspiration-driven strategy evolution~(\ref{rule}), since cooperators are well surrounded by extortioners, whom themselves can obtain positive profits when confronting $C$, much better than getting nothing when confronting $D$. As such, by reasonable aspirations, defectors can even be driven out in the population~[Figs.~\ref{ecd}(a)-(c)]. Contrarily, the checkerboard-like ordering of $E_{\chi}$ and $C$ can be easily destroyed by defectors provided myopic best response rule is used for strategy updating, since their presence in place of an extortioner may yield a higher payoff in a predominantly cooperative neighborhood. Thus, unlike the three-strategy competing game driven by myopic best response rule, where extortion serves as ``the Trojan Horse" and helps cooperators to grab a hold among defectors~\cite{Szolnoki2014pre}, extortion plays the role of ``the Aegis" to impede the invasion of defection, supports the growth and guarantees the flourish of cooperators in our aspiration-driven game model.

\section{Conclusions}
To summarize, we have studied spatial evolutionary Prisoner's dilemma game with and without extortion by adopting aspiration-driven strategy updating rule. Using both Monte Carlo simulations and the generalized mean-field approximations we have determined how the fraction of cooperation depends on the presence of extortion strategy and on the scheme of strategy updating. In games entailing only two strategies $C$ and $D$, we find that asynchronous strategy updating manner always inhibits the evolution of cooperation, in stark contrast to the case of synchronous updating, where cooperation is promoted in a way resembling a coherence-resonance-like behavior. Remarkably, introducing extortioners into the population will wipe off the difference induced by different strategy updating manners, and cooperation is always enhanced for small and moderate aspiration levels. For appropriate parameters, defectors can even be eliminated completely in the population [Figs.~\ref{ecd}(a)-(c)]. Despite of the success of cooperators, another noteworthy property is that extortion itself is always evolutionary stable in our proposed game model with aspiration-driven strategy updating rule. Just like the (myopic) best response rule, the stochastic learning rule based on individual aspiration~(\ref{rule}) is arguably an integral part of human behavior~\cite{Traulsen2010pnas}, our presented results therefore may help us to understand, in addition to the evolution of cooperation, the evolution of extortion in our human society.

\acknowledgments{}
This work was supported by the National Natural Science Foundation of China (Grant No. 11135001), and by the Fundamental Research Funds for the Central Universities (Grant No. lzujbky-2014-28).

\end{document}